\documentclass[aps,prd,twocolumn,groupedaddress,showpacs]{revtex4}
\usepackage{epsf,epsfig,graphicx}
\def\edth{\;\raise1.0pt\hbox{$'$}\hskip-6pt\partial\;}
\def\baredth{\;\overline{\raise1.0pt\hbox{$'$}\hskip-6pt
\partial}\;}
\def\gsim{~\rlap{$>$}{\lower 1.0ex\hbox{$\sim$}}}

\newcommand{\be}{\begin{equation}}
\newcommand{\ba}{\begin{eqnarray}}
\newcommand{\ee}{\end{equation}}
\newcommand{\ea}{\end{eqnarray}}

\newcommand{\fr}{\frac}
\newcommand{\tF}{\tilde{F}}

\begin{document}

\title{Imprint of scalar dark energy on cosmic microwave background polarization}

\author{Seokcheon Lee$^{1,2}$, Guo-Chin Liu$^{3}$, and Kin-Wang Ng$^{1,4}$}

\affiliation{
$^1$Institute of Physics, Academia Sinica, Taipei 11529, Taiwan\\
$^2$Korea Institute for Advanced Study, Seoul 130-722, Korea\\
$^3$Department of Physics, Tamkang University, Tamsui, New Taipei City 25137, Taiwan\\
$^4$Institute of Astronomy and Astrophysics, Academia Sinica, Taipei 11529, Taiwan
}

\vspace*{0.6 cm}
\date{\today}
\vspace*{1.2 cm}

\begin{abstract}
We study the imprint of a coupling of scalar dark energy to a photon
on the cosmic microwave background polarization. Both the time-evolving field value and the perturbation
of the scalar induce $B$-mode polarization. For a wide range of scalar dark energy models allowed by current
observational data, we conclude that future cosmic microwave background data will find either a cosmic parity violation in a temperature-polarization correlation
due to the field value, or perturbation-induced $B$-mode polarization that is indistinguishable from that generated by
primordial gravitational waves.
\end{abstract}

\pacs{95.36.+x, 98.70.Vc}
\maketitle

The existence of a smooth and inert component with an effective negative
pressure, dubbed the dark energy, supported by several observations particularly for the Hubble
diagram of type-Ia supernovae (see, for example,
Refs.~\cite{Bahcall, Riess}), is a big mystery in cosmology;
perhaps its existence is a clue for us to unveil the cosmic codes.
To understand the microscopic nature as well as to measure cosmological signals of dark energy
have become among the most important goals in cosmological research.

The negative pressure of dark energy that accelerates the expansion of the present Universe
is akin to that proposed in the inflationary scenario for resolving so-called
cosmological problems~\cite{olive}. Inflation models usually utilize one or multiple scalar
fields that roll slowly down a scalar potential. The scalar potential is a source of
vacuum energy whose negative pressure drives an exponential expansion of the early Universe.
By the same token, it is natural for considering the scalar field as a candidate
for dark energy and the present Universe is indeed coasting to another
stage of inflation.  Many scalar dark energy models  have been proposed to
account for the present acceleration of the Universe and further constrained by
observational data~\cite{DErev}.

Recently, the discovery of a new particle at the Large Hadron Collider~\cite{LHC},
which is compatible with the properties of the Standard Model Higgs boson,
have alluded to the existence of fundamental scalars as well as the realization of
the Higgs mechanism for symmetry breaking. This has also given
a foundation to considering scalar field as the fundamental theory for inflation or dark energy.
In fact many attempts have been made to model inflation or dark energy as a symmetry breaking
process, in which the vacuum expectation value (vev) of the scalar field plays the role as an
order parameter in a phase transition from a state of higher vacuum energy to the ground state~\cite{olive}.

In this article, we consider an axionlike field model for dark energy in which
the axionic or pseudoscalar $\Phi\equiv M\phi$ couples to the electromagnetic field strength via
$(-\beta/4)\phi F_{\mu\nu} \tF^{\mu \nu}$, where $\beta$ is a coupling
constant and $M$ is the reduced Planck mass.  In this model, the mean field
or the vev of $\phi$ breaks parity symmetry. (Instead, $\phi$ can be
treated as a scalar that has such a parity-violating interaction.)
Recently, it was proposed that string theory suggests the presence of a plenitude of axions
(an axiverse), possibly populating each decade of mass down to the Hubble scale,
and naturally coupled to photons~\cite{arvanitaki}. Models of string axions as candidates for
dark energy and their cosmological constraints have been discussed~\cite{panda}.

It is known that the above $\phi$-photon interaction may lead to
cosmic birefringence~\cite{carroll} that induces the rotation of the polarization plane of the cosmic
microwave background (CMB), thus converting $E$-mode into $B$-mode polarization~\cite{lue,LLN}.
In certain dark energy models, the time-evolving mean field generates rotation-induced $B$-mode polarization with
parity-violating $TB$ and $EB$ correlations, in addition to the standard $TE$ correlation
($T$ represents CMB temperature anisotropy)~\cite{LLN}. Recently, constraints on cosmic parity violation
have been derived from CMB $TB$ and $EB$ measurements~\cite{wmap9}.
Here we will consider the contribution of $\phi$ perturbation to cosmic birefringence~\cite{Li,Caldwell,pospelov}.
We will find that the nature of dark energy may be probed by ongoing CMB experiments or the
cosmic birefringence should be taken into account in interpreting CMB $B$-mode polarization data.

We assume a conformally flat metric,
$ds^2=a^2(\eta) (d\eta^2- d \vec{x}^2)$,
where $a(\eta)$ is the cosmic scale factor and $\eta$ is the conformal time
defined by $dt=a(\eta)d\eta$.
The $\phi F \tF$ term leads to a rotational velocity of the polarization plane of a photon
propagating in the direction $\hat{n}$~\cite{carroll},
\be
\omega(\eta, \vec{x})=-{\beta\over2}\left(\fr{\partial \phi}{\partial\eta}+
\vec{\nabla} \phi \cdot \hat{n}\right).
\label{rot}
\ee
Thomson scatterings of anisotropic CMB photons by free electrons
give rise to linear polarization, which can be described
by the Stokes parameters $Q(\eta, \vec{x})$ and $U(\eta, \vec{x})$.
The time evolution of the linear polarization is governed by the collisional Boltzmann equation,
which would be modified due to the rotational velocity of the polarization plane~(\ref{rot}) by including
a temporal rate of change of the Stokes parameters:
\be
\dot Q\pm i \dot U = \mp i2\omega \left( Q\pm i U \right),
\label{QUeq}
\ee
where the dot denotes $d/d\eta$.
This gives a convolution of the Fourier modes of the Stokes parameters
with the spectral rotation that can be easily incorporated into the Boltzmann code.

Now we consider the time evolution of $\phi$.
Decompose $\phi$ into the vev and the perturbation:
$\phi(\eta,\vec{x}) =\bar{\phi}(\eta) + \delta \phi (\eta, \vec{x})$.
For the metric perturbation, we adopt the synchronous gauge:
$ds^2= a^2(\eta) \{d\eta^2- [\delta_{ij}+h_{ij}(\eta, {\vec x})]dx^i dx^j\}$.
Neglecting the backreaction of the interaction, we obtain the mean-field evolution as
\be
\ddot{\bar\phi} + 2 {\cal H} \dot{\bar\phi} +
\fr{a^2}{M^2} \fr{\partial V}{\partial \bar\phi}= 0\, ,\label{phi0eq}
\ee
where ${\cal H} \equiv \dot{a} / a$ and $V(\phi)$ is the scalar potential.
The equation of motion for the Fourier mode $\delta\phi_{\vec k}$ is given by
\be
\ddot{\delta\phi}_{\vec k}
+ 2 {\cal H} \dot{\delta\phi}_{\vec k} + \left( k^2+
\fr{a^2}{M^2} \fr{\partial^2 V}{\partial \bar\phi^2}\right) {\delta\phi}_{\vec k}=
-{1\over2} {\dot h_{\vec k}} {\dot{\bar\phi}}\, .
\label{fourierphi}
\ee
where $h_{\vec k}$ is the Fourier transform of the trace of $h_{ij}$.

To proceed the calculation, we need to specify the initial conditions for the perturbation
${\delta\phi}_{\vec k}(\eta)$ at an early time $\eta_i$. Let us consider
the means and fluctuations of the energy density
and pressure of $\phi$, given, respectively, by
\ba
&&\bar\rho_{\phi}=\frac{M^2}{2a^2} \dot{\bar\phi}^2 + V(\bar\phi)\, ,\quad
\bar p_{\phi} =\frac{M^2}{2a^2} \dot{\bar\phi}^2 - V(\bar\phi)\, ,\nonumber \\
&&\delta\rho_{\phi}(\vec k,\eta)=\frac{M^2}{a^2} \dot{\bar\phi}\,\dot{\delta\phi_{\vec k}} +
\frac{\partial V}{\partial{\bar\phi}} \delta\phi_{\vec k}\, ,\nonumber \\
&&\delta p_{\phi}(\vec k,\eta)=\frac{M^2}{a^2} \dot{\bar\phi}\,\dot{\delta\phi_{\vec k}} -
\frac{\partial V}{\partial{\bar\phi}} \delta\phi_{\vec k}\, .
\ea
Inflation creates a nearly scale-invariant primordial power spectrum
of adiabatic density perturbations in all light fields. It means that
the entropy perturbation for the entire fluid, just after inflation, vanishes:
$T \delta s = \delta p - (\dot{\bar p}/\dot{\bar\rho}) \delta \rho = 0$.
This enables us to find the relation between $\dot{\delta\phi}$ and $\delta\phi$.
From the fact that long-wavelength fluctuation modes are frozen outside the horizon,
we also set $\dot{\delta\phi}(\eta_i)= 0$.

In Ref.~\cite{dave}, the authors considered two initial conditions, the so-called smooth
and adiabatic cases. The former is that $\dot{\delta\phi}=\delta\phi= 0$; the latter is
that $\delta\rho_\phi/{\bar\rho_\phi}=\delta\rho_r/{\bar\rho_r}=(4/3)\delta\rho_m/{\bar\rho_m}$,
where the last two are the radiation and matter energy densities respectively.
The adiabatic case has this form because the equation of state of $\phi$ equals to that of radiation
($w = w_{r} = 1/3$) in the early radiation-dominated epoch.
It was shown that the CMB anisotropy power spectrum is insensitive to the choice of initial conditions.
The difference in the anisotropy power spectrum is much smaller than the cosmic variance limit in both cases.
We can use the initial condition, $\delta\rho_\phi/{\bar\rho_\phi} \simeq 10^{-16}$.
This implies that $\delta\phi(\eta_i)\simeq 10^{-16}$ and $\dot{\delta\phi}(\eta_i)
= 0$. It was also shown that the isocurvature initial condition can be ignored
because the isocurvature perturbation decays with time~\cite{LLN}.
Thus, we can just consider the adiabatic initial condition for our calculation.

If $\phi$ is nearly massless or its effective mass is less than the present Hubble parameter,
the mass term and the source term in Eq.~(\ref{fourierphi}) can be neglected.
In this case, the dark energy behaves just like a cosmological constant
with $\dot{\bar\phi}=0$. However, its perturbation is dispersive and can be cast into
${\delta\phi}_{\vec k}(\eta)={\delta\phi}_{\vec k,i}\,f(k\eta)$,
where ${\delta\phi}_{\vec k,i}$ is the initial perturbation amplitude
and $f(k\eta)$ is a dispersion factor.
For a superhorizon mode with $k\eta\ll 1$, $f(k\eta)=1$; the factor then
oscillates with a decaying envelope once the mode enters the horizon.
Let us define the initial power spectrum $P_{\delta\phi}(k)$ by
$\left<\delta\phi_{\vec k,i}\delta\phi_{\vec k',i}\right> = (2\pi^2/k^3)
P_{\delta\phi}(k) \, \delta({\vec k}-{\vec k}')$.

\begin{figure}[htbp]
\centerline{\psfig{file=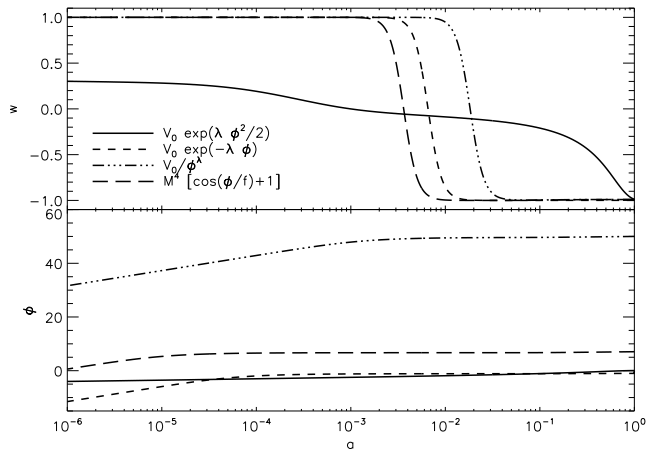, width=9cm}
}\caption{Evolution of the equation of state, $w$, and the mean field of $\phi$ in scalar dark energy models,
whose respective parameters are chosen as to obtain the evolution consistent with current observational data.}
\label{fig:wphi}
\end{figure}

\begin{figure}[htbp]
\centerline{\psfig{file=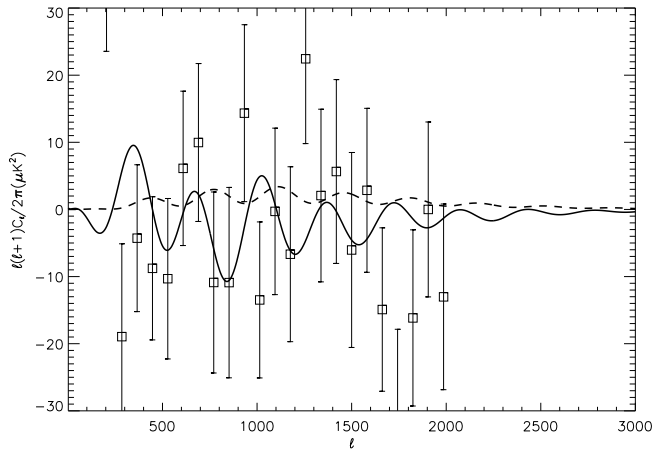, width=9cm}
}\caption{Cosmological birefringence induced $TB$ (solid) and $EB$
  (dashed) mode power spectra. Also shown are the $TB$ measurements
  made by QUaD.}
\label{fig:TBEB}
\end{figure}

\begin{figure}[htbp]
\centerline{\psfig{file=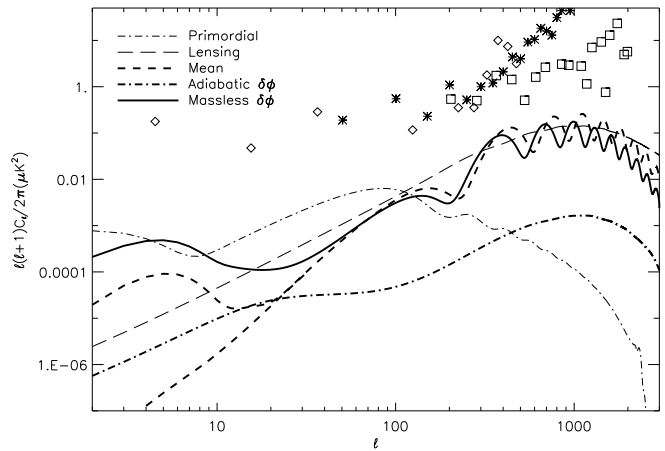, width=9cm}
} \caption{
Cosmological birefringence induced $B$-mode power spectra through the
mean field (thick, dashed), perturbed scalar field with the adiabatic initial condition
(thick, dot-dashed; the power is boosted up with $\beta=10^8$), and
perturbed nearly massless scalar field (thick, solid). For the mean-field spectrum, at low $l$ it branches
into two curves: the spectrum with a peak corresponds to the original scalar model and its power is much suppressed
when enforcing the scalar model with $w=-1$ for $0 \le z \le 10$.
Upper limits ($95 \%$ C.L.) of QUaD (square), Quiet (asterisk), and WMAP9 (diamond) are shown for
comparison. Also shown are the theoretical power spectra of the lensing induced
$B$ mode (thin, dashed) and gravity-wave induced $B$ mode (thin, dot-dashed) with $r=0.13$.}
\label{fig:BMODE}
\end{figure}

In Fig.~\ref{fig:wphi}, we plot the evolution of $w$ and $\bar{\phi}$ in some representative
scalar dark energy models, adjusting each model's parameters such that the evolution is consistent
with current observational data. We have also exploited many other models such as the parametrized
equation of states and more complicated scalar potentials. The obtained results are quite similar to
one another. Here we just provide the results of a typical exponential potential,
$V(\phi)=V_0 \rm{exp}(\lambda \phi^2/2)$, with $\lambda=5$.
We compute the $TB$, $EB$, and $B$-mode power spectra induced
by $\bar{\phi}$ with this potential using our full Boltzmann code based on the CMBFast~\cite{SZ}.
The induced power spectra basically follow the shapes of the $TE$ and $E$-mode
spectra as shown in Figs.~\ref{fig:TBEB} and~\ref{fig:BMODE}.
Also shown are the observed data from WMAP 9-year data~\cite{wmap9},
QUaD, and Quiet~\cite{quad}. We scale the coupling constant $\beta=0.03$ to obtain
the power spectra that are still allowed by the data.
We have tried to modify the scalar model such that the equation of state of scalar dark energy
is forced to be $w=-1$ for $0 \le z \le 10$.
This stops the evolution of $\bar{\phi}$ after the reionization epoch,
thus removing the power on large angular scales, as shown by the absence of
a reionization peak of the mean-field induced $B$-mode power spectrum
in Fig.~\ref{fig:BMODE}. In fact, this suppresses some power on
small angular scales too, so we have used a larger value of
$\beta=0.33$ to produce the $B$-mode power spectrum.
Interestingly, the large-scale $B$ polarization can be used to measure
the equation of state of dark energy through the birefringence effect.

To obtain the induced $B$-mode power spectrum
due to the perturbation $\delta\phi$
for both adiabatic and smooth initial conditions,  we run our Boltzmann
code to compute $\delta\phi_{\vec k}(\eta)$.
The obtained $B$-mode power spectrum is tiny, with a peak about $10^{-31} \mu K^2$
for $\beta=1.0$, and insensitive to initial conditions.
We have tried a modified scalar model in which $w=1$ for $2 \le z \le 20$.
Even though the overall power is increased by 3 orders of magnitude,
it is still well beyond any current experimental sensitivity.
In Fig.~\ref{fig:BMODE}, we plot the power spectrum for the adiabatic initial condition,
which has been boosted by taking $\beta=10^8$ in order to show the spectral feature.
The high-$l$ peak of the power spectrum comes from
the polarization generated on the decoupling surface while the
low-$l$ peak comes from the reionization surface.

For the case that $\phi$ is nearly massless, we solve for $f(k\eta)$ numerically
using Eq.~(\ref{fourierphi}) with $\dot{\bar\phi}=0$ and the initial power spectrum
$P_{\delta\phi} (k)=Ak^{n-1}$, where $A$ is a constant amplitude squared and $n$ is the spectral index.
The space-time background has no difference from that of the Lambda cold dark matter model.
Assuming a scale-invariant spectrum ($n=1$) and $A\beta^2=0.01$,
the induced $B$-mode polarization is computed and plotted in Fig.~\ref{fig:BMODE}.
Our $B$-mode power spectrum is qualitatively consistent with that obtained
in Ref.~\cite{pospelov}, in which models of new particle physics containing
massless pseudoscalar fields superweakly coupled to photons have been considered.
Note that both $C_l^{TB}$ and $C_l^{EB}$ power spectra, unlike the homogeneous case, vanish.
We have also produced the rotation power spectrum~\cite{Li,Caldwell},
\be
C_l^{\alpha}=\frac{\beta^2}{2\pi}
\int {dk}{k^2} \left\{ {\delta\phi}_k (\eta_s)\, j_l[k(\eta_0-\eta_s)]\right\}^2\,,
\label{clalpha}
\ee
where $\eta_s$ denotes the time when the primary CMB polarization is generated
on the last scattering surface or the rescattering surface.
We find that the rotation power spectrum for the recombination case is roughly
$l^2 C_l^{\alpha}\sim 0.005$ for $l<100$ (also see Ref.~\cite{Caldwell})
and $l^4 C_l^{\alpha}\sim 3$ for $l>100$.
For reionization $l^3 C_l^{\alpha}\sim 0.008$ for $l<10$ and $l^{4.5} C_l^{\alpha}\sim 0.01$ for $l>10$.
Recently, constraints on direction-dependent cosmological birefringence from WMAP
7-year data have been derived, with an upper limit on the quadrupole of a scale-invariant
rotation power spectrum, $C_2^{\alpha}<3.8\times 10^{-3}$~\cite{Gluscevic}.
Our quadrupole is within this limit. In fact, the limit should become weaker for
our case because our $C_l^{\alpha}$ scales as $l^{-4}$ for $l>100$.

There have been physical constraints on $A$ and $\beta$. Let us assume that
inflation generates the initial condition for dark energy perturbation.
Then, $n\simeq 1$ and $A \simeq (H/2\pi)^2/M^2$, where $H$
is the Hubble scale of inflation. The recent CMB anisotropy measured by
the Planck mission has put an upper limit on $A< 3.4\times 10^{-11}$~\cite{planck}.
This implies that the present spectral energy density of
dark energy perturbation relative to the critical energy density,
$\Omega_{\delta\phi}< 10^{-15}$, which is negligible compared
to that of radiation. The most stringent limit on $\beta$ comes from
the absence of a $\gamma$-ray burst in coincidence with
Supernova 1987A neutrinos, which would have been converted in the galactic magnetic field
from a burst of axionlike particles due to the Primakoff production in the supernova
core: $\beta < 2.4\times 10^7$ for $m_\phi < 10^{-9}{\rm eV}$~\cite{sn1987}.
Hence the combined limit is $A\beta^2<2\times 10^4$, which is much bigger
than the value that we have used here.

As long as dark energy is birefringent, there would be a wide window for us to see its
properties through measurements of the CMB polarization.
Dynamical dark energy would rotate $E$ polarization into $B$ polarization, thus leaving
cosmic parity-violating $TB$ and $EB$ correlation as well as rotation-induced $B$ modes.
These $B$ mode power spectra are similar to the lensing $B$ mode and the gravity-wave induced $B$ mode,
so the parity violation is crucial to distinguishing between them. Even though dark energy is
indeed a cosmological constant, its perturbation can still generate a rotation-induced $B$ mode
power spectrum while conserving the cosmic parity. In this case, it is a big challenge to do
the separation of different $B$ mode signals. It is apparent that the rotation-induced
$B$ mode has acoustic oscillations but to detect them will require next-generation experiments.
In principle, one may use de-lensing methods~\cite{delensing}
or lensing contributions to CMB bi-spectra~\cite{bispectra} to single out the lensing $B$ mode.
Furthermore, de-rotation techniques can be used to remove the rotation-induced
$B$ mode~\cite{kamion09}. More investigations along this line should be done before
we can confirm the detection of the genuine $B$ mode.

This work was supported in part by the National Science Council, Taiwan, ROC under Grants
No. NSC101-2112-M-001-010-MY3 (K.W.N.) and No. NSC100-2112-M-032-001-MY3
(G.C.L.).

\end{document}